\documentstyle{mn}
\title[Quasi-toroidal oscillations]
{Quasi-toroidal oscillations \\
in rotating relativistic stars
}
\author[   Y. Kojima   ]
{    Yasufumi Kojima \\
 Department of Physics, Hiroshima University,  \\
         Higashi-Hiroshima 739, Japan              }
\date{Accepted ?
      Received ?;
      in original form ?}
\pubyear{199?}  
\begin{document}
\maketitle

\begin{abstract}
Quasi-toroidal oscillations in slowly rotating stars are examined
in the framework of general relativity.
The oscillation frequency
to first order of the rotation rate
is not a single value even for 
uniform rotation unlike the Newtonian case.
All the oscillation frequencies of the r-modes 
are purely neutral and 
form a continuous spectrum limited to a certain range.
The allowed frequencies are determined by 
the resonance condition between the perturbation and
background mean flow. 
The resonant frequency varies with the radius
according to general relativistic dragging effect.
\end{abstract}

\begin{keywords}
oscillation, neutron star
\end{keywords}

\section{ INTRODUCTION  }

In recent X-ray observation with 
the Rossi X-Ray Timing Explorer (RXTE),
quasi-periodic oscillations are discovered in 
several sources (e.g., van der Klis et al.  1996 and 
the subsequent papers of the volume). 
The frequency ranges from a few Hz to kHz and may be 
attributed to the phenomena near a compact object. 
Several models are proposed as the oscillations:
beat-frequency between a magnetized neutron star 
and accretion disc, stellar oscillation, and so on.
For example, 
Strohmayer and Lee (1996) considered the excitation of 
the g- and r-modes as a result of the thermonuclear flash 
and discussed the observational possibility.
Their calculations are however based on the Newtonian 
gravity.   Unlike the spheroidal modes like f-, p-, and g-modes,
the general relativistic effects are not clear for the r-mode,
since the problem has never been studied so far.
The toroidal motion is trivial in a non-rotating star,
but has non-vanishing frequency in a rotating star.
The quasi-toroidal mode is called r-mode and known as 
the Rossby wave in ocean.
Papaloizou and Pringle \shortcite{papa78} introduced the r-mode 
in connection with the variable white dwarfs.
See also the subsequent study 
(Provost, Berthomieu \& Rocca 1981; Saio 1982).

In this  paper, we will explore the relativistic effect 
on the r-mode.
We never discuss the observational implication of
the r-mode in relativistic stars, but theoretical study of 
the oscillation frequency may be a useful tool 
for the future observation.
We use the slow rotation approximation and
linearized Einstein equations. The first-order 
effect of the rotation rate is taken into account.
In section 2, we present the perturbation equations 
describing the r-mode.
In section 3,  the eigenvalue problem is solved.
Finally, section 4 is devoted to the discussion.
Throughout this paper we will use units of $G = c = 1$. 
 
\section{Perturbation equations }

We assume a star with a uniform angular velocity 
$ \Omega \sim O( \varepsilon ) $, and consider the rotational 
effect of order $ \varepsilon $ only.
The configuration of the pressure $ p $ and the density $ \rho $ 
is the same as in the non-rotating star,
since the centrifugal force deforming the shape is
of the order  $ \varepsilon ^2 . $
The metric for the slowly rotating star is given by 
\cite{hartle}
\begin{eqnarray}
 ds^2 & = & -e^\nu dt^2   + e^\lambda  dr^2
           + r^2 ( d \theta ^2 + \sin ^2 \theta d \phi ^2 )
                    \nonumber \\
      & & - 2 \omega r^2 \sin ^2 \theta dt d \phi ,
\end{eqnarray}
where  $ \omega   \sim O( \varepsilon  ) $
is a radial function describing the dragging of the inertial frame.
Introducing  a function $ \varpi = \Omega -\omega $,
we have a differential equation as
\begin{equation}
\left( j r^4 \varpi ' \right)' -
16 \pi (\rho +p ) e^{ \lambda } j r^4 \varpi =0,
\label{eqnw2}\end{equation}
where a prime means a derivative with respect to $r$, and
\begin{equation}
 j = e^{ -( \lambda + \nu ) /2 } .
\end{equation}
The function $ \varpi $ inside the star is 
monotonically increasing function of $r $, so that the range is
limited to 
\begin{equation}
  \varpi _0 \leq    \varpi \leq   \varpi _R , 
\label{eqnwrng}\end{equation}
where $ \varpi_0 $ and $ \varpi_R $ are the values at the center
and surface ($r=R$), respectively.

The perturbations describing non-radial oscillations with the small 
amplitude can be given by the density perturbation $\delta \rho$,
pressure perturbation $\delta p$, and three components of the 
velocity $(U, R, V)$.
The metric perturbations can be expressed by the ten functions,
but the number is reduced to six 
$(h_0, h_1, H_0, H_1, H_2, K)$
by the gauge fixing.
We here use the same notation for these perturbation functions 
as in Kojima \shortcite{kojima92}, but the explicit forms 
are not necessary for most of the following discussion.
In this way, the equations governing the oscillations
are one thermo-dynamic relation and ten components of 
the linearized Einstein equations for these eleven functions. 
In the case of non-rotation, two sets are completely decoupled. 
One set ($ U , h_0, h_1 $) is called 
axial perturbation (or ``odd-parity'' mode), 
while the other set 
($\delta \rho, \delta p, R, V, H_0, H_1, H_2, K $)
polar perturbation (or ``even-parity'' mode). 
Notice that the axial perturbation $ U $ describes the toroidal 
motion, and has zero frequency in the non-rotating star
(Thorne \& Campolattro  1967).
We expect that with rotation the toroidal oscillations of the fluids
have finite frequencies of the order of $ \Omega $  
like in the Newtonian pulsation theory 
(Papaloizou \& Pringle 1978; 
Provost, Berthomieu \& Rocca 1981; Saio 1982).
There also exists gravitational wave mode with non-vanishing
frequency in the axial perturbation
(Chandrasekhar \& Ferrari 1991; Kokkotas 1994). 
The mode can be distinguished from the r-mode in the non-rotating
limit. We never discuss the gravitational wave mode any more here.

We look for the r-mode oscillations in the relativistic rotating stars.
The perturbation functions are expanded by appropriate sets of 
spherical harmonics with index  $l, m $ and
$ \exp [- i (\sigma t -m \phi )] $.
The linearized Einstein equations in the slowly rotating
star are schematically given by
\begin{equation}
{\cal A}_{lm} + {\cal E}_A \times {\tilde {\cal P}} _{l\pm 1m} =0,
\end{equation}
\begin{equation}
{\cal P}_{lm} + {\cal E}_P \times {\tilde {\cal A}} _{l\pm 1m} =0,
\label{eqnpol}
\end{equation}
where ${\cal A} $ and $ {\tilde {\cal A}} $ 
represent some sets of the axial perturbation functions,
while ${\cal P }  $ and ${\tilde {\cal P}}$ represent
those of the polar perturbation functions  \cite{kojima92}. 
The symbols,  ${\cal E}_P $ and $ {\cal E}_A $ are
some operators of the order $ \varepsilon $.
It is clear that the presence of the rotation  
induces  the couplings between the axial and polar modes.
The coupling  is subject to  the selection rule:
the axial mode with  $ l, m $
is coupled with the polar modes with $ l \pm 1 , m$
and vice versa.
This rule  is easily understood if we notice that
the slow rotation perturbation corresponds to the odd-parity
perturbation with $ l=1 $ (Campolattro  \& Thorne 1970).

In the previous papers (Kojima 1992, 1993a,b),
the pulsation equations in the slowly rotating stars
are  examined, assuming that
the oscillation frequencies in the non-rotating stars 
are regarded as non-zero values.
This is true for the spheroidal modes like 
f-, p-, g-modes and gravitational wave modes.
The eigenvalue problems are solved for the non-rotating stars, 
and the rotational corrections are calculated for 
these oscillation modes.
We instead assume that 
the frequency $ \sigma $ is of the order of $ \varepsilon $.
Different manipulation is therefore necessary,
since the rotation should be included at the lowest order 
to obtain the oscillation frequency of the r-mode.
The perturbation functions should be ordered 
in the magnitude as,
\begin{equation}
   \begin{array}{lll}
 h_0 \sim O(U), & h_1 \sim O(\varepsilon^1 U), & \\ 
 \delta \rho \sim O(\varepsilon^1 U), &
 \delta p \sim O(\varepsilon^1 U),    &    \\
  H_0 \sim O(\varepsilon^1 U), &
  H_2 \sim O(\varepsilon^1 U), & 
   K \sim O(\varepsilon^1 U),  \\
 H_1 \sim O(\varepsilon^2 U), &
 R \sim  O(\varepsilon^2 U),  &
 V \sim O(\varepsilon^2 U) .
     \end{array}
\end{equation}
The velocity  perturbations and the metric perturbations with  
$ h_{tj} $ are  'anti-symmetric' with respect to time and  
others like density perturbations are 'symmetric'.  
The former should therefore have even power of $ \varepsilon $, 
while the latter odd power of $ \varepsilon $.
The polar perturbation functions should be  
of higher order in the quasi-toroidal oscillations.
From the above ordering, the linearized Einstein equations 
correspond to those for the axial part, 
${\cal A}_{lm}  =0 $ at the lowest order.
The polar part is induced through the coupling, 
equation 
(\ref{eqnpol}) at higher order level. 
The corrections to the axial part are also induced at 
higher order level.
In proceeding to the higher order,
the higher order rotational corrections for the equilibrium 
states are necessary, but the corrections of the order 
$ \varepsilon $ are sufficient at the lowest order.

We now solve $ {\cal A}_{lm} =0 $ for $ U , h_0, h_1 $. 
The quasi-toroidal velocity can be expressed as 
\begin{equation}
  \left( \sigma - m \Omega +{2m \varpi \over l(l+1)} 
\right) U 
 =  - 4 \pi ( \sigma - m \Omega ) 
 (\rho +p) e^{- \nu}  h_0 .
\label{eqnuh}\end{equation}
The relation  between the metric perturbations is  
\begin{equation}
h_1 = - {  i r^4 e^{-\nu} \over  (l-1)(l+2) } 
\left[ ( \sigma - m \Omega) \Phi '  
    +{2m \omega ' \over l(l+1)} \Phi 
\right],
\end{equation}
where
\begin{equation}
 \Phi = { h_0 \over r^2 }.
\end{equation}
The master equation governing the quasi-toroidal 
oscillations can be written as
\begin{equation}
 \left( \varpi -\mu \right)
 \left[  {1 \over  j r^4 } \left( j r^4 \Phi ' \right) '
  - v \Phi  \right]  =  q \Phi ,
\label{eqnbase1}\end{equation}
where 
\begin{equation}
  v =  { e^{ \lambda } \over r^2 }
\left[ l(l+1) -2 \right] , 
\end{equation}
\begin{eqnarray}
q &=& {1 \over  j r^4 } \left( j r^4 \varpi ' \right)' 
\\
&=& 16 \pi (\rho +p ) e^{ \lambda } \varpi ,
\label{eqndefq}\end{eqnarray}
and  the eigenvalue
\begin{equation}
  \mu =  - { l(l+1) \over 2m  } ( \sigma -m \Omega) .
\end{equation}
In equation (\ref{eqndefq}), we have used the relation (\ref{eqnw2}).

\section{  SINGULAR EIGENVALUE PROBLEM }
The basic equation (\ref{eqnbase1}) is not a regular eigenvalue problem.
The coefficient $ ( \varpi -\mu ) $ becomes singular inside 
the star for a certain value of $\mu $.  
The coefficient also vanishes outside the star, but 
the singularity can be removable because $ q = 0$.
This equation is very analogous to the Rayleigh's equation 
for the incompressible shear flow (e.g., Lin 1955). 
The perturbation propagating with the wave number $k$ and speed $c$
in the mean flow with velocity $u$
can be described as
\begin{equation}
 \left(  u-c \right) \left[ \Phi '' -k^2 \Phi \right] = u'' \Phi .
\end{equation}
The similar singular eigenvalue problems appear in many other fields, 
e.g., differential rotating fluid discs and plasma oscillations. 
See e.g., Balmforth \& Morrison  \shortcite{bal95} 
for the methods of solving the singular eigenvalue problem.
The singular point is called as critical layer 
in the fluid dynamics,
or co-rotating point in the rotating discs.
The studies of the singular eigenvalue problem indicate that
unless there is an inflection point,
$ u '' =0 $, somewhere within the flow,
the eigenvalue is not discrete, but continuous and neutral 
against the stability.
All neutral modes must have critical layers (co-rotation points)
that lie within the flow, and therefore form a continuous 
spectrum of intrinsically irregular eigen-functions.

The parallel argument holds for our problem.
The essential points are that the potential 
$ v $ is  positive definite for $ l>2$
and that there is no inflection point ($ q >0 $) inside the stars.
We can conclude that the eigenvalue of 
equation (\ref{eqnbase1}) is real number and
the range is limited to
\begin{equation}
  \varpi _0 <   \mu =
- {l(l+1) \over 2m} ( \sigma -m \Omega) 
    <   \varpi _R , 
\label{eqnrang}\end{equation}
where the range of $\varpi $, (\ref{eqnwrng}) is used.

\medskip
We shall simply show the conclusion by 
{\it reductio ad absurdum.}  
If there is a non-trivial solution of which 
the eigenvalue $ \mu $ is not located within the domain
(\ref{eqnrang}), then we have the integral relation
\begin{eqnarray}
0&=&  \int _0 ^\infty  \left( 
   \left| \Phi ' \right|^2  + v \left| \Phi \right|^2 \right)
    j r^4 dr
\nonumber \\
&&+ \int _0 ^R  { 1 \over \varpi -\mu }
    q | \Phi |^2   j r^4 dr, 
\label{eqnint1}\end{eqnarray}
where we have assumed that the function $\Phi$ tends to zero
both at the center and the infinity. 
The imaginary part of equation (\ref{eqnint1}) gives 
\begin{equation}
0 = \Im (\mu) \int _0 ^R  { 1 \over |\varpi -\mu |^2 }
q | \Phi |^2   j r^4 dr. 
\end{equation}
Since $q$ is positive definite for $ 0 \leq r < R $ as seen in 
equation (\ref{eqndefq}), we have
$ \Im (\mu)  = 0$, except the trivial case $ \Phi =0 $.
That is, the eigenvalue $ \mu $ must be real number.
In a similar way, we introduce $ \Phi =  ( \varpi -\mu) f $
to have another integral relation,
\begin{equation}
0=  \int _0 ^\infty  \left( \varpi -\mu \right)^2
  \left( \left| f ' \right|^2  + v \left| f \right|^2 \right)
j r^4  dr .
\label{eqnint2}\end{equation}
The function within the integral is positive at least 
for $ 0 \leq r < R $. We therefore have the contradiction.
\medskip

If the eigenvalue is located within the domain (\ref{eqnrang}),
the eigen-function has the singular point, say, $r_*$, 
inside the star.
The function is approximated by the delta-function as
$ f \sim \delta(r-r_*) $.
The quasi-toroidal fluid velocity has the  
form $ U \sim \delta ( r-r_*) $ from the equation (\ref{eqnuh}).
The function represents a steep resonance 
between the perturbation and the mean flow.
The resonance may be more clear,
if we consider so-called the Cowling approximation.
In the Newtonian pulsation theory,  
gravitational perturbations are sometimes neglected 
in the oscillations. This trick gives good results for
the spheroidal mode as well as the r-modes.
The relativistic Cowling approximation is given by
$ \delta T^{\alpha \beta} _{\ \ \ ; \beta} =0 $ with  
$ \delta g_{\alpha \beta} =0 $.
One component of the equations is reduced in the 
slow rotation case to
\begin{equation}
 ( \varpi -   \mu ) U  = 0 .
\end{equation}
It is clear that the solution of this equation is 
$ U \sim \delta ( r-r_*) $, and that 
the range of eigenvalue is given by equation (\ref{eqnrang}).

Finally, we consider the Newtonian limit, 
in which $ \varpi  \to \Omega $. The frequency 
therefore corresponds to a single value as
\begin{equation}
\sigma = \left( 1 - {2  \over l(l+1) }  \right) m\Omega .
\end{equation}
This is the frequency of the r-mode oscillation
measured by inertial frame.

\section{  DISCUSSION }

In this paper, the r-mode oscillation is examined as the 
consistent first-order solution to the quasi-toroidal motion.
The frequency forms a continuous spectrum.
The oscillation is caused by a certain resonance 
between the perturbation and the background rotating flow.
The resonance condition is that the co-rotating frequency, 
$ (\sigma - m \Omega ) e^{ -\nu /2} $  of the wave 
should be $ - 2m /(l(l+1)) $ times the angular velocity,
$ \varpi e^{ -\nu /2} $ 
measured by ZAMO (zero-angular-momentum-observer). 
The angular velocity depends on the position of 
the local inertia frame due to the dragging effect.
In this way, the r-mode oscillations in relativistic 
stars are much analogous to those in the differential rotating 
discs, although the angular velocity, $\Omega $ is uniform.
The mechanism works everywhere within the rotating star,
but the resultant frequency, $\sigma $ measured at infinity 
is not identical. 
This is the physical meaning of the continuous spectrum of 
the r-mode.

The eigen-function of the Newtonian r-mode is not determined 
to first-order of the rotation,
since any functions for the same $ \mu $
satisfy the equation governing the oscillation,
\begin{equation}
 (\mu - \Omega ) U  = 0 .
\end{equation}
The modes are degenerated in this sense.  In order to
determine the radial structure of the r-modes,
calculation of  the next order is necessary. 
The higher order corrections to the frequency 
will remove the degeneracy.
As for the relativistic r-mode, the frequencies are 
distinguished corresponding to the resonance points.
All the positions are on an equal footing
to the first-order of the rotation.
Therefore, the normal frequency forms a  continuous spectrum.
We expect that some favored resonance points are selected
as a result of the higher order corrections.
That is, the axial part of the first order
drives the density and pressure perturbations at the
second order.
The gravitational radiation may be also 
associated with the density perturbations. 
The polar perturbations react on the frequency at the third order.
The internal structure will strongly affect the modes through
the coupling.
The relevant second-order rotational corrections
$ \sim O(\varepsilon ^2 ) $  like rotational deformation 
are of course necessary to solve the problem.
The study beyond the first-order of the rotation
is very important not only for the radial structure, 
but also for the stability,
although the calculations are significantly complicated.

The frequency at the first order is a real number, and 
the mode represents standing ripple in the rotating flow. 
The wave will decay or grow due to the dissipation. 
The gravitational radiation reaction and/or
the viscosity cause the instability of spheroidal modes 
in the rotating star.
Similar instability may set in for the r-mode, according to 
the general argument (Friedman \& Schutz 1978;
Friedman \& Morsink 1997).
Recent numerical calculation suggested the instability of 
the r-mode (Andersson 1997). 
However, these works are not in agreement as for
the growth rate, which is higher order consequence of 
$\varepsilon ^n ( n \ge 2) $.

In conclusion,
the second-order effect to the r-mode oscillation 
in the relativistic star is  complicated, 
but quite interesting problem.

\section*{  ACKNOWLEDGMENT }

   I would like to acknowledge the hospitality of
the Albert Einstein Institute in Potsdam 
where most of this work was done. 
I also thank N. Andersson, K.D. Kokkotas, 
and B.F. Schutz for their discussion.
This was supported in part
by the Grant-in-Aid for Scientific Research Fund of
the Ministry of Education, Science and Culture of Japan 
(08640378).
%


\end{document}